\documentstyle[12pt]{article}
\advance\textheight by \topskip
\newcommand{\BE}{\begin{equation}}
\newcommand{\EE}{\end{equation}}
\newcommand{\BA}{\begin{eqnarray}}
\newcommand{\EA}{\end{eqnarray}}

\def\e{\varepsilon}
\def\s{\sigma}
\def\ad{\alpha ' }
\def\B{\beta}
\def\p{\partial}
\def\no{\nonumber}
\def\bi{\bibitem}
\def\vp{\varphi}
\def\tb{\bar{t}}
\def\rb{\bar{r}}
\def\th{\theta}
\def\a{\alpha}
\def\at{\tilde{\a}}
\def\Tb{\bar{T}}
\def\am{\alpha_{m} }
\def\atm{\tilde{\alpha}_{m} }
\def\an{\alpha_{n} }
\def\atn{{\tilde{\alpha}}_{n} }

\def\bra#1{\left\langle #1\right|}
\def\ket#1{\left| #1\right\rangle}

\input epsf

\begin{document}

\rightline{May 97}
\rightline{OU-HET 266}
\rightline{hep-th/9705100}

\vspace{.8cm}
\begin{center}
{\large\bf The Information Loss Problem of Black Hole and the First Order Phase Transition in String Theory}

\vskip .9 cm

{\bf Kenji Hotta,}
\footnote{E-mail address: hotta@funpth.phys.sci.osaka-u.ac.jp}

Department of Physics, \break
Graduate School of Science, Osaka University, \break
Toyonaka, Osaka 560, JAPAN
\vskip.6cm

\end{center}
\vskip .6 cm
\centerline{\bf ABSTRACT}
\vspace{-0.7cm}
\begin{quotation}

In recent years, Susskind, Thorlacius and Uglum have proposed a model for strings near a black hole horizon in order to represent the quantum mechanical entropy of the black hole and to resolve the information loss problem. However, this model is insufficient because they did not consider the metric modification due to massive strings and did not explain how to carry information from inside of the horizon to the outside world. In this paper, we present a possible, intuitive model for the time development of a black hole in order to solve the information loss problem. In this model, we assume that a first order phase transition occurs near the Hagedorn temperature and the string gas changes to hypothetical matter with vanishing entropy and energy which we call `the Planck solid'. We also study the background geometry of black holes in this picture and find out that there is no singularity within the model.

\end{quotation}

\normalsize
\newpage

\section{Introduction}
\label{sec:Intro}

Imagine that we are observing the gravitational collapse of a star to a black hole from a distance. According to general relativity, the motion of the constituents of the star should look extremely slow because of time dilation in the strong gravitational field. When would the black hole be formed for us? According to Hawking's theorem, the black hole would radiate out like a black body with finite temperature \cite{Haw}. What is the mechanism of the  Hawking radiation? All the matter falls into the singularity, and nothing is known afterward. Then, where has the information gone? In this paper, we give one possible resolution to these naive questions.

The information loss problem stated above is one of the most serious difficulties in present physics \cite{Pre}. The essence of this problem is the fact that the quantum state of Hawking radiation does not depend on the initial state of the collapsing body. This is because the state of the radiation is determined only by the outside geometry of the black hole horizon. Let us suppose that the Hilbert space \( \cal H \) is factorized into \( {\cal H}_{in} \) and \( {\cal H}_{out} \), which correspond to the inner and outer sides of the event horizon, respectively. Taking the states \( \ket{\psi_{in} (t)} \in {\cal H}_{in} \) and \( \ket{\psi_{out} (t)} \in {\cal H}_{out} \), we can represent the state \( \ket{\psi (t)} \) of the entire space as
\BE
  \ket{\psi (t)} = \ket{\psi_{in} (t)} \otimes \ket{\psi_{out} (t)}.
\EE
The information of the initial state of the collapsing body is included in \( \ket{\psi_{in} (t)} \), since \( \ket{\psi_{out} (t)} \), which consists of the Hawking radiation state, does not depend on the initial state. This implies that the information is lost forever, because no matter can escape from the black hole. In other words, since there is no correlation between the initial state \( \ket{\psi_{initial}} \) and the final state \( \ket{\psi_{final}} \) which appears when the Hawking radiation has finished, we cannot trace the time development of the quantum states of the black hole from the information of the collapsing body as long as we remain within the framework of the usual quantum theory and general relativity. This fact implies that the unitarity of quantum theory does not hold.

A related question arises in quantum statistical mechanics. If the black hole has entropy, it must be able to be represented by counting the number of quantum states. 't Hooft has shown that if one introduces a spatial cutoff at about a Planck length separated from the event horizon, one can represent the Bekenstein-Hawking entropy by computing the entropy of scalar fields outside of the black hole \cite{tht1}. This implies that the information may be stored outside the horizon.

In recent years, Susskind, Thorlacius and Uglum have constructed a model for strings near the black hole horizon in order to resolve the problem of information loss by representing the Bekenstein-Hawking entropy quantum mechanically \cite{sus1} \( \sim \) \cite{sus6}. They proved that for an observer far from the black hole, strings near the event horizon are thermalized and stored between the event horizon and the stretched horizon, the latter of which is positioned a string-scale length away from the former. They also proved that, in particular when the black hole mass is infinite, we can represent the black hole entropy by counting the number of string states. These facts suggest that information regarding the matter falling into the black hole is stored in the long strings near the horizon. Let us call this model the `stretched horizon model'. However, there are two problems in this model. We must consider the mechanism under which information is carried outside of the horizon and the metric modification by massive strings. We will see details of these problems in \S \ref{sec:Stretched}, together with a review of the stretched horizon model. 

We wish to reconstruct this model while keeping its advantages. In \S \ref{sec:Time}, we show that if we assume the existence of matter with vanishing entropy and energy, which we call `the Planck solid', we can construct a scenario in which the information loss problem does not exist from the viewpoint of an outside observer. In this scenario, the information is excluded from the inside of the black hole due to the character of the Planck solid, and it is stored in the string gas near the horizon, as in the stretched horizon model. This string gas provides almost all the contribution to the mass energy of the black hole. Then, we propose a toy model, which is later referred to as the `string bit model', that gives us an intuitive picture of the Planck solid in \S \ref{sec:Stringbit}. It is concluded that as the temperature grows, the string gas in the weak coupling region changes to the Planck solid in the strong coupling region by a first order phase transition. A first order phase transition in string theory has been proposed by Sathiapalan \cite{Sa}, Kogan \cite{Ko}, Atick and Witten \cite{AW}. This is seen in \S \ref{sec:Stringbit}.

With regard to the problem of metric modification, we introduce a new metric to treat string condensation in a spherical shell region around the Planck horizon, which is defined as the boundary of the Planck solid, in \S \ref{sec:Background} \cite{shell2}. Because the solution has no singularity due to the vanishing energy of the Planck solid, we need not think that there exists a singularity in a black hole. In \S \ref{sec:entropy} we argue that our model does not contradict the Susskind-Uglum calculation of the entropy \cite{sus5}. Finally, a summary and discussion are presented in \S \ref{sec:Summary}.

\section{The stretched horizon model and its problems}
\label{sec:Stretched}

We first review the stretched horizon model. For simplicity, we use general relativity for the background metric and treat the energy momentum tensor of matter classically in the entire this paper. We adopt the natural units \( c= \hbar = k_{B} =1 \), but use \( G \) for the gravitational coupling constant.

Susskind and Thorlacius performed a gedanken experiment concerning observation around the black hole based on an analogy between the Schwarzschild space-time and the Rindler space-time, and proposed the black hole complementarity principle \cite{sus3}. They concluded that an observer at infinity would see high temperature phenomena near the event horizon. It is natural to think that a freely falling object is thermalized by the gravitational effect. Susskind has considered what would be observed at infinity when strings are falling into the black hole \cite{sus4}. Let us assume that the background field is spherically symmetric, and take the Schwarzschild metric. Its line element is given by
\BE
  ds^{2} = \left( 1 - \frac{2GM}{r} \right) dt^{2}
          - \left( 1 - \frac{2GM}{r} \right)^{-1} dr^{2}
           - r^{2} d \Omega_{2}^{2},
\label{eq:schmetric}
\EE
where \( M \) is the black hole mass and
\BE
  d \Omega_{2}^{2} = d \th^{2} + \sin^{2} \th \ d \vp^{2}.
\EE
Let us consider, for example, a string falling freely straight into the black hole, leaving from a stationary observer. It seems for this observer that this string never falls beyond the event horizon, due to the slowing of time in the strong gravitational field. This situation is analogous to that of strings moving at near light velocity against a stationary observer in flat space. We take light-cone coordinates and define \( X_{\perp} ( \s ) \) as the string coordinates in the transverse direction at the world sheet time \( \tau  = 0 \). We can expand this as
\BE
  X_{\perp} ( \s ) = x_{\perp} + \sqrt{\frac{\ad}{2}} \sum_{l} \left( \frac{ \a_{l} }{l} e^{il \s} + \frac{ \at_{l} }{l} e^{-il \s } \right),
\EE
where \( x_{\perp} \) is the center-of-mass coordinate of the string, and \( \a_{l} , \at_{l} \) are the operators of oscillating modes which obey the commutation relations
\BA
  \left[ \am , \an \right] &=& \left[ \atm , \atn \right]
     = m \delta_{m+n,0}, \\
  \left[ \am , \atn \right] &=& 0.
\EA
Even if this string is in the ground state, its mean-square-length in the transverse direction diverges. This results from the infinitely high frequency oscillating modes as
\BE
  R_{\perp}^{2} \equiv \bra{0} [ X_{\perp} ( \s ) - x_{\perp} ]^{2} \ket{0}
    = \frac{\ad}{2} \sum_{l=1}^{\infty} \frac{1}{l}.
\EE
We must renormalize this expression since it diverges logarithmically. The Hamiltonian is given by
\BE
  H = \frac{p_{\perp}^{2} + m^{2}}{2 P_{+}},
\EE
where \( P_{+} \), \( p_{\perp} \) and \( m \) are the longitudinal momentum, the transverse momentum and the mass of strings, respectively. When the resolution time of the stationary observer is \( \e \), a high energy cutoff \( E < 1/ \e \) is introduced into the Hamiltonian. Since the contribution of oscillating modes is included in \( m \), this cutoff corresponds to the high frequency cutoff of the string oscillations for \( l < P_{+} / \e \), if we ignore \( p_{\perp} \). Introducing this high frequency cutoff, Susskind has concluded that
\BE
  R_{\perp}^{2} = \frac{\ad}{2} \sum_{l}^{ \frac{P_{+}}{\e} } \frac{1}{l}
     \simeq \frac{\ad}{2} \ln \frac{P_{+}}{\e}.
\label{eq:lengthperp}
\EE
A similar calculation can be performed for the mean-square-length in the longitudinal direction, and is given by
\BE
  R_{+}^{2} \equiv \bra{0} [ X^{+} ( \s ) - x^{+} ]^{2} \ket{0}
    \simeq \frac{\ad}{2} \left( \frac{P_{+}}{\e} \right)^{2}.
\label{eq:length-}
\EE
Thus, both quantities increase as \( P_{+} \) grows. If we define the proper time of the stationary observer \( \tau_{o} \) and that of the string \( \tau_{s} \), the longitudinal momentum \( P_{+} \) behaves as
\BE
  P_{+} \simeq 2m \left( \frac{d \tau_{s}}{d \tau_{o}} \right)^{-1}.
\label{eq:longiprop}
\EE
The time \( \tau_{s} \) elapses very slowly because the string is moving at near the light velocity. Therefore, the string is oscillating very slowly, even if it has high frequency modes. This is the reason for the movement of the cutoff frequency. 

Let us now return to the freely falling string situation. Since \( P_{+} \) behaves as
\BE
  P_{+} \sim \exp \left( \frac{t}{2GM} \right),
\label{eq:longifree}
\EE
near the horizon, the transverse size \( R_{\perp} \) increases as \( t^{1/2} \). In this case, the relation between the proper time of the freely falling string, \( \tau \), and that of the observer at infinity, \( t \), is approximated near the event horizon as
\BE
  \frac{d \tau}{dt} \sim \exp \left( - \frac{t}{2GM} \right).
\label{eq:proptfree}
\EE
The time corresponding to the string is frozen, as observed by the outside observer, since \( d \tau / dt \) decays exponentially with \( t \). Because Eqs. (\ref{eq:longifree}) and (\ref{eq:proptfree}) satisfy Eq. (\ref{eq:longiprop}), the length of the string grows due to the slowing of their proper time. Let us call this phenomenon `the thermalization effect'. If such a process is repeated many times, then large strings are stocked near the horizon. The surface where the stringy nature appears is the so-called stretched horizon \cite{sus1}. We can see the high energy phenomena of strings in this region.

At the next step, Susskind and Uglum computed the entropy of an infinite mass black hole using string theory provided that strings are in an equilibrium state around the event horizon \cite{sus5}. The equilibrium state in the stationary curved background space implies that the temperature \( \Tb \) measured by the stationary observer at each point satisfies \cite{LL}
\BE
  \sqrt{ g_{00} } \ \ \Tb = const,
\label{eq:temequil}
\EE
where \( g_{00} \) is the time component of the metric tensor there. In our Schwarzschild black hole case, for \( g_{00} =1 \) at infinity, the constant above is the temperature \( T \) at infinity. If we identify the temperature \( T \) as the Hawking temperature \( T_{BH} = 1/8 \pi GM \), then the temperature \( \Tb \) is given by
\BE
  \Tb = \frac{ 1 }{ 8 \pi GM \left( 1 - \frac{2GM}{r} \right)^{1/2} }.
\label{eq:teminequill}
\EE
This implies that the temperature is higher at points closer to the horizon. Hence we can observe the high temperature physics near the horizon. In the case of an infinite mass black hole, they concluded that the entropy \( \s \) of the strings in a unit area on the horizon becomes
\BE
  \s = \frac{1}{4G},
\EE
which agrees with the Bekenstein-Hawking entropy \( S_{BH} \) when it is multiplied by the area \( A \) of the event horizon; namely,
\BE
   A \s = S_{BH} \equiv \frac{A}{4G}.
\EE
This means that the strings near the horizon has all the degrees of freedom of the black hole, since the thermal entropy gives the maximal number of states of a system. Therefore, it is natural to conclude that all the information of the collapsing body existing before remains in the strings near the horizon.

Unfortunately, although the stretched horizon model is very successful, there are two problems: 

\begin{enumerate}

\item 
\label{it:prob1}
The black hole is assumed to exist from the outset, and only strings falling into it after its formation are discussed in the model. However, if we wish to resolve the information loss problem, we must consider information corresponding to the matter which existed before its formation.

\item 
\label{it:prob2}
There are extremely long strings near the event horizon as seen by the observer at infinity, but these have large mass coming from the oscillating mode energies and affect the gravitational field. Therefore, we must correct the metric at the interior of the black hole.

\end{enumerate}

The first problem above is related to the question of what we will see when a black hole is formed. In particular, we are interested in where the information concerning the collapsing matter has gone. Thus, we must study the time development of the black hole. We will mention a possible resolution in the next section. 

For the second problem above, if we assume, for simplicity, that the energy density of strings has an upper limit at the Planck energy density \( \rho_{p} \) which appears at the Planck temperature \( T_{p} \), we can easily calculate the mass of the strings between the surface at \( \Tb = T_{p} \) and the event horizon. From the above discussion, the radius \( r_{p} \) at the Planck temperature is obtained as
\BE
  T_{p} = \frac{ 1 }{ 8 \pi GM \left( 1 - \frac{2GM}{r_{p}} \right)^{1/2} }.
\EE
When the black hole mass \( M \) is sufficiently large, the radius \( r_{p} \) is nearly equal to the Schwarzschild radius \( r_{s} \equiv 2GM \), so that we can represent it as \( r_{p} = 2GM + \delta \) with an infinitesimal variable \( \delta \) (\( \delta >0 \)). From the above equation and \( T_{p} = G^{-1/2} \), \( \delta \) is given by
\BE
  \delta \simeq \frac{1}{ 32 \pi^{2} M}.
\EE
In our spherically symmetric case, we can obtain the mass \( m \) from the simple formula
\BE
  m = \int 4 \pi r^{2} \rho dr,
\label{eq:schmass}
\EE
where \( \rho \) is the energy density (note that this integration is not over the proper volume of the Schwarzschild metric \cite{Sz}). The mass of the strings between the surface at \( T = T_{p} \) and the event horizon is given by
\BE
  m = \int_{2GM}^{2GM+ \delta } 4 \pi r^{2} \rho_{p} dr \sim M,
\EE
where we have approximated this integration up to first order in \( \delta \) and have used \( \rho_{p} = G^{-2} \) (the fact that the coordinate volume is proportional to \( M \) plays an important role). Because this is comparable to the black hole mass \( M \), it is expected that there is little mass inside the black hole and the metric is very smooth, unlike the Schwarzschild metric.

This may appear strange, because, to this time, the coordinates of the black hole have been taken as the Schwarzschild metric over the entire space, including the inside of the horizon. But, we are convinced of this for two reasons. First, based on the general principle of relativity, physical law must be the same whatever coordinates we use. If massive strings are detected by an observer at infinity, they must generate a gravitational interaction. Secondly, even if the metric of the inside of the black hole is the Schwarzschild metric, it is not observed by outside observers. Of course, this does not mean that background metric always becomes an observable one. As we will see, however, we can construct a metric which is always observable by outside observers.

\section{The time development of the black hole}
\label{sec:Time}

Now, let us consider the time development of the black hole when we see it from outside. In general relativity, the surface of a collapsing body never disappears from our sight, because the proper time is elongated, and if it reaches the Schwarzschild radius, then the time is frozen forever against the outside observer. Thus, although radiation from it is extremely red-shifted, it can be observed eventually. Similarly, we can see all the matter of the collapsing body as well (of course, we cannot see the inside). We have not succeeded in formulating strings in curved space-time, including their quantum effect and the back reaction to the background metric. But, however complex the physical phenomena, the physical property of strings, which cutoff frequencies increase due to the dilation of the proper time, does not change. Thus, even if  Eqs. (\ref{eq:lengthperp}) and (\ref{eq:length-}) are not valid, string thermalization occurs in the strong gravitational field. When the factor \( d \tau / dt \) vanishes, strings are thermalized at any temperature in so far as no new physical phenomena occur. It is expected that we are able to see high temperature phenomena in the collapsing body.

In order to connect this evolution of the black hole to its Hawking radiation, we assume that the matter of the collapsing body will finally reach a quasi-equilibrium state at the Hawking temperature. To satisfy this condition, the matter must be thermalized enough to support the gravitational force by its pressure, and the temperature of the inner matter must be larger than that of outer matter. This corresponds to the argument of the notion of the thermal equilibrium in curved space discussed in the previous section. If we naively construct the scenario for black hole formation from the gravitational collapse, it is inferred as follows:

\begin{enumerate}
\item 
\label{it:proc1}
When the gravitational collapse of a star begins, the energy density of matter becomes larger and its temperature rises. For an observer at infinity, the surface of the star exists outside the Schwarzschild radius, and the matter inside is observable.

\item
\label{it:proc2}
The proper time for the in-falling matter elapses more slowly, and the matter experiences a thermalization effect. Its energy density will be dominated by extremely long strings. These phenomena occur at the center of the star first and spread outward as time goes on.

\item
\label{it:proc3}
By this thermalization effect, the strings obtain a larger energy than in the case without this effect, and when the pressure of the strings becomes sufficiently large to balance the gravitational force of the matter, the strings stop falling. Then this star is in a quasi-equilibrium state at the Hawking temperature, and Hawking radiation begins from its surface.

\end{enumerate}

In process \ref{it:proc2}, the energy of the matter is dominated by that of a few very long strings, based on the thermodynamics of an ideal gas of strings. The nature of this system was investigated (for example, the papers \cite{Tan} and references therein). With regard to process \ref{it:proc3}, note that the energy of the strings is pulled out from the vacuum by the thermalization effect. This system never reaches an equilibrium state because the black hole has negative specific heat. Hawking radiation is interpreted as radiation from high temperature strings.

However, this scenario has two problems, as was pointed out by Susskind and Griffin \cite{sus6}. First, if the matter is distributed in three dimensions, its mass energy is proportional to \( M^{3} \), because the Schwarzschild radius \( r_{s} \) is proportional to \( M \). This conflicts with the fact that the black hole has mass \( M \) and the Schwarzschild metric (\ref{eq:schmetric}). Also, since the entropy is an extensive variable, the entropy \( S \) of the matter is proportional to \( M^{3} \), which differs from the Bekenstein-Hawking entropy \( S_{BH} \propto M^{2} \). These facts show that the matter must not be distributed three dimensionally. Moreover, Susskind and Uglum's entropy arguments cannot be applied to this scenario, because they compute the entropy of the strings spreading two dimensionally on the horizon.

It may appear that our approach fails to represent the black hole state. However, it is sufficient if there is a mechanism for the exclusion of the entropy and the energy from the inside of the black hole. We assume that at high temperature, the string gas will be transformed into matter with a vanishing entropy and energy by a first order phase transition. Let us call this matter `the Planck solid'. Because this matter has no energy, it stays in the same state for all time. At the first sight, this assumption appears strange. But this is not totally unreasonable, as we shall see in the next section. Adopting this assumption, we can construct a new scenario as follows:

\begin{enumerate}
\item 
\label{it:proc1+}
When the gravitational collapse of a star begins, the energy density of the matter becomes larger, and its temperature rises. For an observer at infinity, the surface of the star exists outside the Schwarzschild radius, and the matter inside is observable.

\item
\label{it:proc2+}
The proper time for the in-falling matter elapses more slowly, and the matter experiences a thermalization effect. Its energy density will be dominated by extremely long strings. These phenomena happen at the center of the star first, and then spread outwards as time goes on.

\item 
\label{it:proc3+}
If the energy density of the strings reach a value near the Planck energy density, the phase transition occurs and the Planck solid appears. This phenomenon also first happens in the center of the star, and the Planck solid region extends its size, while causing energy and entropy of the matter to move outwards. Let us call the surface of the Planck solid region `the Planck horizon'.

\item
\label{it:proc4+}
When the Planck horizon reaches outside of the Schwarzschild radius, the gravitational collapse is over, and Hawking radiation begins.

\end{enumerate}
\begin{figure}[ht]
\begin{center}
$${\epsfxsize=10.0 truecm \epsfbox{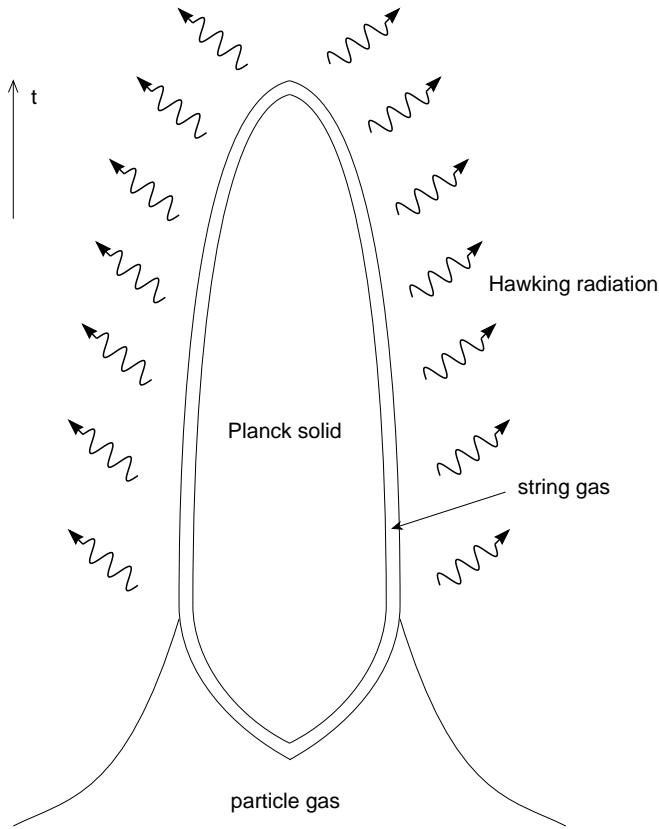}}$$
\caption{The time development of a black hole}
\label{fig:blacktimef}
\end{center}
\end{figure}

In Figure \ref{fig:blacktimef}, we represent the time development of the black hole in this scenario. According to this scenario, the information contained by the star is pushed out from the inside of the black hole by the Planck solid because it can possess no information, and it is stored in the strings outside of the black hole, as in the stretched horizon model! Hawking radiation is interpreted in this model as the radiation from the string gas which contains the information of the collapsing matter. In this process, the Planck solid changes to the string gas at its surface to supplement the lost string gas. If we factorize the Hilbert space into the inner and outer regions of the Planck horizon, then we can represent the total state \( \ket{\psi (t)} \) as
\BE
  \ket{\psi (t)} = \ket{\psi_{in}} \otimes \ket{\psi_{out} (t)}.
\EE
All the evolution occurs outside of the Planck horizon, since \( \ket{\psi_{in}} \) does not depend on time and is fixed in only one state. Therefore, \( \ket{\psi_{out} (t)} \) depends on the initial condition, and contains all the information. This means the state \( \ket{\psi (t)} \) is described by the time development from the initial state. The state \( \ket{\psi (t)} \) is not the thermal ensemble state, and we can deduce information from the difference between the actual radiation and a thermal radiation. Moreover, because there is no energy inside, it is expected that the background metric has no singularity. We discuss the Planck solid in the next section and the background metric in \S \ref{sec:Background}.

\section{The string bit model and the first order phase transition}
\label{sec:Stringbit}

In this section, we propose a toy model for strings in order to obtain an intuitive picture of the phase transition in a high temperature string system. To begin with, let us recall the high temperature ideal gas of strings, namely, the string thermodynamics in the weak coupling region. As stated in the previous section, the energy of strings near the Hagedorn temperature is dominated by the oscillating mode energy, i.e., the mass energy of a single string. This is because the number of oscillating modes increases exponentially with energy, and the probability of the energy domination by a single long string is very high. This is the origin of the upper limit of the temperature, i.e., the Hagedorn temperature, in the ideal gas of strings \cite{Hag}. The entropy \( S \) is expressed as a function of energy \( E \) as \cite{Tan}
\BE
  S \simeq \B_{H} E,
\label{eq:stringent}
\EE
where \( \B_{H} = 1/ T_{H} \) is the inverse of the Hagedorn temperature. If the energy density increases up to the Planck energy density, however, the coupling of strings becomes too large to treat it by a perturbation method, and it seems that the smooth Riemann surface of world sheet breaks down \cite{AW}. Unfortunately, we have not succeeded in analyzing such a high density system of strings yet. Thus, in order to obtain an intuitive picture of this system, let us consider the following toy model, which is motivated by the holographic principle \cite{sus7}.

\begin{figure}[ht]
\begin{center}
$${\epsfxsize=5.0 truecm \epsfbox{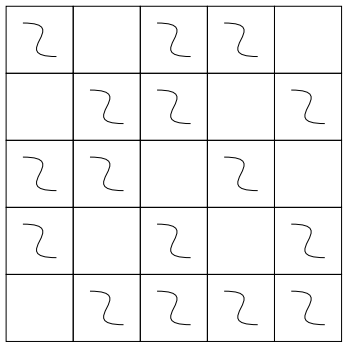}}$$
\caption{The string bit model.}
\label{fig:string_bit}
\end{center}
\end{figure}

We make three assumptions. First, we assume, following Klebanov and Susskind \cite{KL}, that strings consist of `string bits' of Planck length \( l_{p} \) and Planck energy \( \e_{p} \). Similarly, we suppose that space is divided into cubes with Planck-size edges, calling each a `Planckian cell'. Lastly, we assume that we can put only one string bit in one Planckian cell; that is, to each Planckian cell, there are two possible states, that in which it is occupied by a string bit and that in which it is not. In Figure \ref{fig:string_bit}, we represent the string bit model. Wavy lines designate the string bits and boxes the Planckian cells.

Based on these assumptions, let us consider a finite space with \( m \) Planckian cells. We ignore the interactions among the string bits for the time being and treat them as an ideal gas. The total energy \( E \) of this string bit gas system is given by the product of the number of string bits \( n \) and the Planck energy \( \e_{p} \). As there is a one to one correspondence between \( n \) and \( E \), we treat the number of states \( W \) as a function of \( n \) instead of \( E \). If the string bits are undistinguishable, the number of states \( W(n) \) of this system is given by
\BE
  W(n) = {}_mC_n = \frac{m!}{n!(m-n)!}.
\EE
This is the binomial distribution. From this we can see that, if \( m=n \), namely, all the Planckian cells are filled with string bits and the energy density of the system is the Planck energy density, the number of states is \( W(m)=1 \). Therefore, the entropy \( S(n) \) of this system is
\BE
  S(m) = \log W(m) = 0;
\EE
that is, the entropy vanishes. Although a vanishing entropy is preferred as the Planck solid, there is a problem. As \( S \) increases in the small \( n \) region and vanishes at \( n=m \), it must decrease in the large \( n \) region. For the total energy \( E=n \e_{p} \), this means the temperature of this system becomes negative or even divergent, from the formula
\BE
  \frac{1}{T} = \frac{\p S}{\p E}.
\EE
Such a system must not exist, at least as an equilibrium state. But we can avoid this problem by taking into account the coupling of the string bits and a phase transition. 

Here, let us recall the ice-water phase transition. Since molecules of the water move more freely than those of the ice, the entropy of the water \( \s_{water} \) per mole is very much larger than that of the ice \( \s_{ice} \). Thus, this is a first order phase transition with latent heat \( q \) per mole. We have
\BE
  q = T ( \s_{water} - \s_{ice} ),
\EE
from the definition of the entropy. Hence, the water changes the ice by releasing latent heat. Returning to our string bit model, we can take the Planck solid as the ice and the string gas as the water. The string gas with large entropy transforms into the Planck solid with vanishing entropy. Note that the Planck solid is created from the string gas at high temperature (it is the inverse relation of the ice-water case with regard to temperature). From Eq. (\ref{eq:stringent}), the entropy \( \s_{string} \) of this system per string bit is
\BE
  \s_{string} \sim \B_{H} \e_{p}.
\EE
Because the entropy of the Planck solid is zero, the latent heat per string bit released in this phase transition is
\BE
  q \sim T_{H} ( \s_{string} - \s_{Planck} ) \sim \e_{p}.
\EE
This means that all the energy of the string bit is released as latent heat, and hence the Planck solid has no energy. The reason for the vanishing energy of the Planck solid is understood easily by taking into account the interaction of the string bits. The binding energy of the string bits in the Planck solid is the Planck energy, and this energy cancels the mass energy of the string bits. When there are few string bits, they behave as a gas without interaction. But, as the number of the string bits grows, the coupling becomes strong, and this growth does not stop in the region \( m/2 \leq n < m \), because it is energetically favorable to add more string bits, and eventually the Planck solid state comes into existence. To sum up, when the energy density of the string gas approaches the Planck energy density, the interactions change from a weak coupling to a strong coupling, and the first order phase transition occurs. Then the entire space is filled with the strings, and the Planck solid appears by releasing all the entropy and the energy.

It was not our original idea that the first order phase transition in the string system occurs at high temperature. This was first discussed by Sathiapalan \cite{Sa} and Kogan \cite{Ko}. Then, Atick and Witten have proposed the following model based on the relation between string theory and large-N QCD \cite{AW}. They analyzed the effective action and the free energy of the finite temperature system of strings by the Matsubara method. The infinity of the free energy above the Hagedorn temperature \( T_{H} \) comes from the fact that a winding mode in the Euclidean time direction becomes tachyonic \cite{Sa} \cite{Ko}. This implies that the perturbative vacuum state is unstable, so that we must give this mode an expectation value and look for a stable solution. The continuous world sheet picture breaks down in this stable vacuum where the winding mode has its expectation value. Taking into account the contribution of the dilaton, they concluded that a first order phase transition occurs at the critical temperature slightly below \( T_{H} \) with a large latent heat. We know nothing about this new vacuum, but, if we persistently compute the free energy of strings above \( T_{H} \) by a perturbative method, the entropy derived from it becomes smaller than that of the conventional relativistic field theory. This means that there is a lower density of gauge invariant degrees of freedom, which agrees with the idea that there are fewer degrees of freedom inside a distance of the string length scale \( \sqrt{\ad} \) in string theory than in usual field theory. It is expected that string theory is ultraviolet finite, and fewer degrees of freedom are permitted in higher order perturbation than in field theory. In this sense, it is not strange that the string system in the high temperature region has a very small entropy.

Moreover, if we take the Planck solid as a tightly binded system of strings, they mutually restrict each other's freedom, like a solid. In other words, they determine states of the others perfectly by filling up the entire space, and the system is confined in this filled up state. Then the phase transition is interpreted as a transition from the perturbative vacuum to a strong coupling vacuum.

The properties of the Planck solid are satisfied by topological field theories \cite{TFT}. These theories are expected to describe the unbroken phase of general covariance, which corresponds to the confinement phase in QCD \cite{TFTW1} \( \sim \) \cite{TFTW3}. These theories are characterized by the observables which depend only on the topological quantities and are independent of any metric. In this theory, the energy-momentum tensor operator is given by BRST commutator,
\BE
  T_{\mu \nu} = \{ Q_{B} , \lambda_{\mu \nu} \},
\EE
for some operator \( \lambda_{\mu \nu} \), where \( Q_{B} \) is the BRST operator. Since BRST exact operators vanish when inserted into a correlation function, the energy is zero. In addition, another essential property of topological field theories is the absence of dynamical excitation. In other words, there are no propagating degrees of freedom. Thus, matter which is described by the topological field theories has no energy and no entropy, as expected for the Planck solid. In other words, the Planck solid is the unbroken phase of string theory \cite{TFTW3} or the `confinement phase of strings'. 't Hooft suggested that if we can treat the inside of a black hole by topological field theory, all physical degrees of freedom can be projected onto the horizon \cite{tht2}.

\section{The background metric of black hole for the observer at infinity}
\label{sec:Background}

As we have seen, the string gas exists between the stretched horizon and the Planck horizon and generates most of the mass energy of a black hole. To see this, let us make a rough estimation. Suppose that the energy density of the string gas is characterized by the string mass scale, \( \rho_{s} \sim {\ad}^{-2} \), and that the proper distance between the two horizons is characterized by the fundamental string length \( l_{s} \sim {\ad}^{1/2} \). If we represent the radial coordinate \( r \) as \( r = 2GM + \delta \) with the infinitesimal variable \( \delta \) (\( \delta >0 \)), as in the \S \ref{sec:Stretched}, the difference in \( \delta \) between the two horizons is given by
\BE
  \delta_{s} \sim \frac{ \ad }{GM},
\EE
where we have used the formula of the proper distance from the Schwarzschild radius,
\BE
  l = \sqrt{ 8GM (r-2GM)}.
\EE
From this, the coordinate volume of the string gas region is approximated by
\BE
  V_{s} \sim 4 \pi (2GM)^{2} \delta_{s} \sim G \ad M,
\EE
and we can obtain the mass of the string gas system by multiplying the volume \( V_{s} \) by the energy density \( \rho_{s} \):
\BE
  m \sim G {\ad}^{-1} M.
\EE
In particular, when the string scale is almost the same as the Planck scale, \( m \) is  comparable to the black hole mass. Since the proper distance between these two horizons is about the fundamental string length \( l_{s} \), we can ignore this in comparison with the size of the black hole. Thus, we assume that the thickness of this string gas region is zero and impose the spherically symmetric condition on the black hole. The background metric of this spherical thin shell was first discussed by Israel many years ago \cite{shell1}. But this work was based on very complicated Gauss-Codazzi formalism. Here, we refer to the work of Khorrami and Mansouri \cite{shell2} , which is based on an easier distribution method, and discuss the metric of black hole for the observer at infinity. In particular, we compute the background metric when the shell is in a stationary state. This is because, for the large mass black hole, the energy of the Hawking radiation per unit time is extremely small compared to the mass energy of the black hole, and we can ignore the time variation of the metric.

We can apply the Birkhoff theorem to our spherically symmetric case \cite{Wein}, because the spaces of the interior and exterior of the shell are in `vacuums'. Since there is an object with mass \( M \) inside, the line element of the exterior space is given by that of the Schwarzschild coordinate,
\BE
  ds_{+}^{2} = e^{ \nu (r) } dt^{2}
              - e^{ - \nu (r) } dr^{2}
               - r^{2} d \Omega_{2}^{2},
\label{eq:extmetric}
\EE
where
\BA
  e^{ \nu (r) } &=& 1 - \frac{2GM}{r}, \\
  d \Omega_{2}^{2} &=& d \th^{2} + \sin^{2} \th d \vp^{2}.
\EA
For the interior, the line element is given by the flat one,
\BE
  ds_{-}^{2} = d \tb^{2} - d \rb^{2}
               - \rb^{2} d \Omega_{2}^{2},
\label{eq:intmetric}
\EE
because there is no mass energy inside the shell. In Eq. (\ref{eq:intmetric}), we have used the connectivity condition with regard to the angle variables; namely, we take the same variables \( \th \) and \( \vp \) as in (\ref{eq:extmetric}). In so far as we are interested in a stationary background, we assume that the radius \( R \) of the shell is constant all time. In order to connect these metrics on the shell, let us represent inner coordinates \( \rb \) and \( \tb \) as the function of outer ones \( r \) and \( t \),
\BA
  \rb &=& a(r,t), \\
  \tb &=& b(r,t).
\EA
Substituting these into Eq. (\ref{eq:intmetric}), we obtain
\BA
  ds_{-}^{2} &=& \left[ \left( \frac{\p b}{\p t} \right)^{2}
              - \left( \frac{\p a}{\p t} \right)^{2} \right] dt^{2}
               - \left[ \frac{\p b}{\p t} \frac{\p b}{\p r}
                - \frac{\p a}{\p t} \frac{\p a}{\p r}  \right] dr dt \no \\
               &&- \left[ \left( \frac{\p a}{\p r} \right)^{2}
                  - \left( \frac{\p b}{\p r} \right)^{2} \right] dr^{2}
                   - a^{2} d \Omega_{2}^{2}.
\label{eq:intmetric2}
\EA
The connectivity conditions are obtained by identifying the metrics in the line elements (\ref{eq:extmetric}) and (\ref{eq:intmetric2}) at \( r=R \),
\BA
  \B_{t}^{2} - \a_{t}^{2} &=& e^{ \nu (R) }, \\
   \B_{t} \B_{r} - \a_{t} \a_{r} &=& 0, \\
  \a_{r}^{2} - \B_{r}^{2} &=& e^{ - \nu (R) }, \\
  a( R , t ) &=& R,
\EA
where
\BE
  \a_{t} = \left. \frac{\p a}{\p t} \right|_{r=R} \ \ , \ \
  \a_{r} = \left. \frac{\p a}{\p r} \right|_{r=R} \ \ , \ \
  \B_{t} = \left. \frac{\p b}{\p t} \right|_{r=R} \ \ , \ \
  \B_{r} = \left. \frac{\p b}{\p r} \right|_{r=R} \ \ , \ \
\EE
and the solutions of these equations are given
\BA
\a_{r} &=& e^{ - \nu (R) / 2 },
    \label{eq:alphar} \\
  \B_{t} &=& e^{ \nu (R) / 2 },
    \label{eq:Betat} \\
  \a_{t} &=& \B_{r} \ \ = \ \ 0.
    \label{eq:atBr}
\EA
Since the relations (\ref{eq:alphar}) \( \sim \) (\ref{eq:atBr}) are the only connectivity conditions at \( r = R \), there are arbitrary choices of the function \( a( r , t ) \) and \( b( r , t ) \) in entire region of \( r \) and \( t \). This corresponds to the freedom of the general coordinate transformations. For convenience, we take \( a \) as a function of only \( r \) and \( b \) of only \( t \). Then, these functions must satisfy
\BA
  \frac{\p a}{\p r} &\neq& 0, \\
  \frac{\p b}{\p t} &\neq& 0,
\EA
for the finiteness of the inverse metric tensor. For the condition \( a(0) = 0 \) and \( a(R) = R \), \( a(r) \) must be a monotonically increasing function of \( r \). For example, we can choose it as a cubic function of \( r \) as
\BE
  a(r) = r \left[ \a r^{2} + R^{-1} (e^{ - \nu (R) / 2 } - 2 \a R^{2} -1) r
         + ( \a R^{2} - e^{ - \nu (R) / 2 } +2) \right],
\EE
where \( \a \) is a constant and satisfies
\BE
  \a > \frac{ e^{ - \nu (R) / 2 } - 2 }{ R^{2} }.
\EE
On the other hand, from Eqs. (\ref{eq:Betat}) and (\ref{eq:atBr}), \( b( r , t ) \) is linear in \( t \) at \( r = R \); namely,
\BE
  b( R , t ) = e^{ \nu (R) / 2 } t.
\EE
For example, we can choose \( \tb \) as
\BE
  \tb = e^{ \nu (R) / 2 } t;
  \label{eq:enRt}
\EE
that is, time passes in the same way at every point in the interior for the outside observer. Because the string shell is near the Schwarzschild radius \( r_{s} \) (\( R > r_{s} \)), the coefficient \( e^{ \nu (R) / 2 } \) in Eq. (\ref{eq:enRt}) is extremely small. This means that the proper time in the Planck solid region passes much more slowly than that at infinity, and high temperature phenomena are seen in the black hole. In (\ref{eq:intmetric}) let us use the time variable \( t \) defined in (\ref{eq:enRt}) instead of \( \tb \). Then we get the line element
\BE
  ds_{-}^{2} = e^{ \nu (R) } d t^{2} - d \rb^{2}
               - \rb^{2} d \Omega_{2}^{2}.
\label{eq:intslow}
\EE
This implies a flat space, but the time passes extremely slowly. Therefore, in order to take this line element self-consistently, it is sufficient for us to prove the existence of the matter with vanishing energy in the high temperature region only in the flat space.

If we assume that the Planck solid is described by topological field theories as mentioned in the previous section, the observables are independent of the choice of the metric. However, we must perform the gauge fixing to the metric inside to compute the topological quantities. The above connectivity conditions correspond to one of the reasonable gauge fixing conditions which connect the metric inside and outside.

In order to maintain a stationary shell, there must be a force which supports it against its gravitational potential. As we saw above, we can choose the line element of the Planck solid region as (\ref{eq:intmetric}), and the metric is \( \tb \)-independent. The Planck solid changes size only as a result of the phase transition on its surface, just like a solid. In other words, the Planck horizon has a very high pressure, the degeneracy pressure of the Planck solid, since there exists only one quantum state which has been already dominated. If we drop the matter to the black hole in order to squeeze the Planck horizon against its high pressure, it changes into Planck matter, only to extend the Planck solid region. As a result, the Schwarzschild radius and the mass of the black hole increase. This pressure can be taken as a negative tension or a positive pressure in the shell. In our stationary case, the surface tension \( \xi \) is evaluated by Khorrami and Mansouri \cite{shell2} as
\BE
  \xi = - \frac{G \varrho^{2} R}{4(2-G \varrho R)},
\EE
where \( \varrho \) is the surface energy density. Because the shell exists near the Schwarzschild radius \( R \simeq r_{s} \) and the surface energy density is given by
\BE
  \varrho \simeq \frac{M}{4 \pi (2GM)^{2}} = \frac{1}{16 \pi G^{2} M},
\EE
the surface tension takes the very small negative value
\BE
  \xi \simeq - \frac{1}{1024 \pi^{2} G^{2} M}.
\EE
Thus, the shell behaves as if it has this small negative tension.

Although to this point we have discussed from the viewpoint of the observer at infinity, it seems that no observer sees a singularity according to the general covariance. But this is in conflict with the principle of black hole complementarity, which was proposed by Susskind, Thorlacius and Uglum \cite{sus1} \( \sim \) \cite{sus3}. This complementarity principle is summarized in Ref. \cite{sus2} as follows:

\begin{enumerate}

\item To a freely falling observer, matter falling toward a black hole encounters nothing out of the ordinary upon crossing the horizon. All quantum
information contained in the initial matter passes freely to the interior of the black hole.

\item To an observer outside the black hole, matter, upon reaching the ``stretched horizon'', is disrupted and emitted as thermalized radiation
before crossing the horizon.  All quantum information contained in the initial matter is found in the emitted radiation.

\end{enumerate}

This seems somewhat strange at the following point. For example, let us consider 
a rocket freely falling toward the black hole and two observers. One is the observer at infinity and the other is the observer in the rocket. If the freely falling observer has a thermometer, he sees it does not detect the increase in temperature since he receives no Hawking particles based on the above complementarity principle. On the other hand, the observer at infinity sees that the thermometer indicates an increase in temperature due to the thermalization effect. If the observer in the rocket starts the engine before it reaches the horizon and returns back to infinity, the two observers find a discrepancy in their observational results. These facts clearly violate causality (of course the observer in the rocket receives thermal radiation when the engine starts, but this fact has nothing to do with above arguments). In order to preserve causality, at least at the classical level, all the observers must see the same phenomena for the same event at the same point, even if their physical interpretations are different.

We do not know the complete answer, but it seems that the following argument is more reasonable. Susskind, Thorlacius and Uglum deduced this complementarity principle from the relation between the Hawking effect in an infinite mass black hole and the Unruh effect in the Rindler coordinates \cite{Unruh}. The Unruh effect is that by which a constantly accelerating observer receives thermal radiation. In this case, the energy of the thermal particles are supplied by the external accelerating force \cite{BD}. In other words, the thermal particles are created by this accelerating force. At first sight, it seems that the Hawking radiation of the infinite mass black hole corresponds to this Unruh radiation. However, the infinite mass black hole has no Hawking radiation, since it has zero temperature. In this case, an observer fixed at any given point sees the radiation made by the force to maintain their position against the gravitational force from the black hole, as in the Unruh effect. But, this radiation is not that from the black hole and does not contain information about the black hole. In the finite mass black hole case, observers at infinity need not keep their position against the gravitational force, but see the Hawking radiation. Thus, the Hawking radiation is not radiation caused by the external accelerating force. We cannot explain the Hawking effect only by the motion of observers.

Let us return to the first of Hawking's arguments for Hawking radiation \cite{Haw} ( see also Ref. \cite{BD}). If we consider the time development of the Schwarzschild black hole, the space-time can be treated asymptotically as a Minkowski space-time in the remote past, which we refer to as the `in region', and Schwarzschild space-time in the remote future, which we refer to as the `out region'. In the in region, we can choose the `natural' Minkowskian vacuum. We can also choose the `natural' Minkowskian vacuum in the asymptotic region \( r \rightarrow \infty \) in the out region, since the Schwarzschild metric is asymptotically the Minkowski metric. If we compute the wave function using the Schr\"odinger equation in curved space-time, we can define the vacuum states in the two regions. The vacuum state in the in region is different from that in the out region, but they are related by the Bogoliubov transformation. Thus, even if we choose the vacuum state in the in region as an initial state, we obtain as the final state a thermal state which is calculated from the Bogoliubov coefficient of this transformation. From this, we can conclude the creation of Hawking particles. On the other hand, there is no Hawking radiation if the metric has no time dependence, since the vacuum state does not change, and no excitation of particles occurs. If we take causality into consideration, these facts imply that the Hawking particles have been `created' by the time-dependent gravitational field and have already been produced when the metric becomes the Schwarzschild one.

From this viewpoint one may say that the Hawking radiation is real radiation from the black hole, and all the observers see it independent of their states of motion, instead of the complementarity principle. Thus, a freely falling observer receives intensive Hawking radiation as he approaches the Schwarzschild radius, and sees the Planck solid as real matter. We can therefore say that all the observers find out about the creation of the Planck solid and that no one sees singularities, including naked ones, as long as causality at the classical level is maintained. In \S \ref{sec:Time}, we assumed that the string thermalization is caused by the increase of the cutoff frequencies for the observer at infinity. But rather, in order to preserve causality at the classical level, we must confirm that the strings are excited by the time variation of the metric, taking into consideration the quantum effect of strings in curved space-time. In particular, the strings should be thermalized sufficiently for the phase transition before the event horizon appears. It is suggestive that, in Hawking's arguments, the Hawking particles are highly excited at the outside of the event horizon. In any case, we assume that strings are excited and become the Planck solid before the factor \( d \tau / dt \) vanishes, as in the argument of ultraviolet cutoff.

\section{The entropy of the black hole with the Planck solid}
\label{sec:entropy}

In the stretched horizon model, the black hole entropy is computed by counting the states of the strings near the event horizon with the weak coupling assumption \cite{sus5}. But, in our model, we introduce the Planck solid which is expected to appear in the strong coupling region. The question now arises. Does our model contradict their entropy argument? The answer is `no', as we see below.

First, we shortly review the Susskind-Uglum computation of the black hole entropy. If we adopt the Matsubara method and compute the entropy of the equilibrium system of the strings around the event horizon in the Schwarzschild metric, it has an infra-red infinity. This is related to the fact that the temperature at infinity is not zero, but rather it is the Hawking temperature from the argument in \S \ref{sec:Stretched}. Moreover, it is terribly complicated to compute the entropy in the Schwarzschild metric. However, if we consider an infinite mass black hole whose Hawking temperature is zero, this kind of divergence does not appear. For the purpose of analyzing an infinite mass black hole, it is convenient to introduce the Rindler coordinates, which resemble the Kruskal-Szekeres coordinates. In this space, it is easier to compute the entropy than in Schwarzschild space. 

The line element of the Schwarzschild coordinates is given by (\ref{eq:schmetric}). If we perform general coordinate transformations there, we obtain
\BA
  \eta &=& \frac{t}{4GM}, \\
  \rho &=& \sqrt{ 8GM (r-2GM)}.
\EA
Then the result is
\BE
  ds^{2}
   = \rho^{2} \left( 1+ \frac{\rho^{2}}{16G^{2} M^{2}} \right)^{-1} d \eta^{2}
    - \left( 1+ \frac{\rho^{2}}{16G^{2} M^{2}} \right) d \rho^{2}
     - 4 G^{2} M^{2} \left( 1+ \frac{\rho^{2}}{16G^{2} M^{2}} \right)^{2}
      d \Omega_{2}^{2},
\label{eq:SchRin}
\EE
where
\BE
  d \Omega_{2}^{2} = d \th^{2} + \sin^{2} \th \ d \vp^{2}.
\EE
Taking the infinite mass limit \( M \rightarrow \infty \), we obtain the line element for the Rindler coordinates,
\BE
  ds^{2} = \rho^{2} d \eta^{2} - d \rho^{2} - (dx^{2})^{2} - (dx^{3})^{2}.
\label{eq:Rindler}
\EE
Here, we have replaced the angle variables \( \th \) and \( \vp \) by the rectangular coordinate variables \( x^{2} \) and \( x^{3} \) with a proper rescaling. The variable \( \rho \) agrees with the proper length from the Schwarzschild radius \( r_{s} \), and \( \rho = 0 \) corresponds to the event horizon. Note that we are able to use this approximation not only for large values of \( M^{2} \), but also for small values of \( \rho \); that is, near the event horizon for the finite mass. These coordinates are familiar for people in analysing constantly accelerated observers in flat space who experience the Unruh effect \cite{Unruh}.

When we perform the Euclidean continuation with respect to the time variable \( \eta \) in the Matsubara method on Rindler space, the coordinates \( \eta \) and \( \rho \) are transformed into those of the cylindrical space, so that we must take the period of Euclidean time as \( 2 \pi \) in order to avoid a conical singularity. Thus, the temperature in Rindler space is given by \( T_{R} = 1/2 \pi \), which is the so-called `Rindler temperature'. From the argument regarding the notion of temperature in curved space, the temperature at each point in this space is given by \( \Tb = 1/2\pi \rho \). In the \( \rho \rightarrow \infty \) limit, it vanishes, as expected from the Hawking temperature of the infinite mass black hole. However, since the entropy \( S \) is derived from the free energy \( F( \B ) \) by the formula
\BE
  S = \B^{2} \frac{\p}{\p \B} F( \B ),
\EE
we must introduce the \( \B \) dependence of the free energy and compute it on the conical singular background.

Susskind and Uglum evaluated the entropy \( \s \) of strings per unit area in the \( x^{2} \)-\( x^{3} \) plane on the Euclidean conical singular background using the canonical ensemble method. This area corresponds to the unit area on the event horizon. They concluded that it is
\BE
  \s = \frac{1}{4G},
\EE
so that the entropy all over the horizon corresponds to the Bekenstein-Hawking entropy. Because they computed \( \s \) by the canonical ensemble method, it does not take into account the behavior of string gas near the Hagedorn temperature, which is derivable from the microcanonical ensemble method \cite{Tan}.

In our model, on the other hand, we introduce the Planck solid and take the entropy inside of the Planck horizon as zero. This introduces a cutoff at the Planck horizon in the entropy computation. Let us determine the position of the cutoff. For simplicity, we ignore the metric modification outside of the Planck horizon. If we take the temperature of the Planck solid as \( T_{ps} \), the Planck horizon exists at
\BE
  \rho = \rho_{ph} \equiv \frac{1}{2 \pi T_{ps}}.
\EE
Using the dimensionless variable \( \a_{ps} \), we define \( T_{ps} \equiv \a_{ps} G^{-1/2} \). Then
\BE
  \rho_{ph} = \frac{G^{1/2}}{2 \pi \a_{ps}}.
\EE
In the Susskind-Uglum argument, they did not consider the Hagedorn temperature. In this case, the Planck energy density appears at the Planck temperature. Therefore, we must take \( \a_{ps} \) to be of order \( 1 \). Then the cutoff exists at \( \rho_{ph} \sim G^{1/2} \), namely, the Planck length. 

The fundamental parameters in the string theory are the slope parameter \( \ad \) and the expectation value of the dilaton \( < \vp > \). The string coupling \( g \) is related to \( < \vp > \) as \( g \sim e^{ < \vp > } \), and the gravitational coupling \( G \) is represented as \( G \sim g^{2} \). In the above entropy arguments, they used perturbation theory, which would describe the entropy in the case that the string coupling is small. Thus, let us consider the small \( < \vp > \) limit with fixed \( \ad \). This corresponds to taking the small limit of \( G \) while fixing \( M \) in our case. Returning to Eq. (\ref{eq:SchRin}), let us take the coupling \( G \) very small, while keeping \( GM \) large. The Planck horizon moves toward \( \rho =0 \), and the computation of the black hole entropy in this limit agrees with that of Susskind-Uglum. Thus, their computation corresponds to that in a very idealized system in the weak coupling region, ignoring metric modification. Their argument does not contradict our model.

Before the Susskind-Uglum computation, 't Hooft calculated the black hole entropy by state counting near the event horizon in a scalar field theory. This is the so-called `brick wall model' \cite{tht1}. This model gives an entropy proportional to the area of the event horizon when we introduce the ultraviolet cutoff at about one Planck length from the event horizon. The Planck horizon tells us why we need this cutoff.

As an effective model for giving the black hole entropy, there exists the membrane model \cite{mem}. In this model, the membrane must lie at a distance from the event horizon to regularize the physical quantities. Our model gives the reason for this regularization. Namely, in the case that the stretched horizon and the Planck horizon sit close together and we can effectively treat stretched horizon as the membrane which is sitting at about the Planck length from the Schwarzschild radius.

Recently, there has been remarkable progress made in understanding extremal and near extremal charged black holes by D-brane technology \cite{Pol}. It was first shown by Strominger and Vafa that the entropy of a five-dimensional extremal black hole is exactly reproduced by the state counting of open strings on D-branes \cite{SV}. Then, many people confirmed the relation between extremal black holes and a certain configuration of D-branes and strings for quite a variety of black holes \cite{BHST}.

The argument of Strominger and Vafa is based on the supersymmetric nonrenormalization theorem. As one increases the string coupling, the system of strings and D-branes in BPS states must become an extremal black hole. We can calculate the number of states of the strings on the D-branes at weak coupling and then extrapolate the result to the black hole in the strong coupling region, because of the nonrenormalization theorem. Using this method, we can reproduce the entropy of extremal black holes exactly by state counting of strings. However, since black holes are non-perturbative objects, we must understand non-perturbative strings and D-branes in order to discuss the transition from the system of strings and D-branes to the black hole \cite{cor}. Thus, this method does not give the picture of the strings and D-branes in the black hole case. In any case, if outside observers are able to obtain information concerning the black hole and there is no lost information, the information must exist outside of the Schwarzschild radius. In this sense, the Planck solid might be a candidate for the system of strings and D-branes in the non-perturbative region, since it gives a mechanism for the exclusion of the information from the inside of the black hole.

On the other hand, we can obtain a more direct understanding of the five-dimensional extremal black hole in the work of Callan and Maldacena \cite{CM}. Superstring theory is well defined on ten-dimensional space-time. In their argument, a five-dimensional space is compactified into a five-dimensional torus \( T^{5} \), and D5-branes wrap around this torus and D1-branes wrap around one of the compact directions. D-branes exist at the origin of the noncompact four-dimensional space in this configuration. Performing the coordinate transformation, we can obtain the five-dimensional extremal charged black hole solution of the five-dimensional Einstein-plus-Maxwell equation:
\BE
  ds^{2} = \left( 1- \frac{r_{e}^{2}}{r^{2}} \right)^{2} dt^{2}
          - \left( 1- \frac{r_{e}^{2}}{r^{2}} \right)^{-2} d r^{2}
           - r^{2} d \Omega_{3}^{2},
\label{eq:Extrem}
\EE
where
\BE
  d \Omega_{3}^{2} = d \chi^{2} 
                  + \sin^{2} \chi (d \th^{2} + \sin^{2} \th \ d \vp^{2}).
\EE
This extremal black hole has zero temperature but non-zero horizon area. In this coordinate transformation, the origin of the original coordinates is transformed to the surface at the Schwarzschild radius \( r = r_{e} \), namely, the event horizon in (\ref{eq:Extrem}). The black hole entropy is reproduced by the state counting of the strings on the D-branes. Callan and Maldacena argued that strings falling into the black hole would actually appear as open strings whose ends move along the horizon to an outside observer, and this means that the horizon behaves as a D-brane. Since there is no `inside of the horizon' in the original coordinates of the D-brane configuration, open strings exist only outside of the horizon. 

In order to see the relation between the Callan-Maldacena model and the Planck solid model, let us consider the positions of the stretched horizon and the Planck horizon based on the argument of the temperature in curved space in \S \ref{sec:Stretched}. From Eq. (\ref{eq:temequil}), we can obtain
\BE
  \left( 1- \frac{r_{e}^{2}}{r^{2}} \right) \Tb = T_{BH},
\EE
where \( \Tb \) is the temperature at each point, and \( T_{BH} \) is the Hawking temperature. Since \( T_{BH} = 0 \), the radii of the stretched horizon and the Planck horizon become \( r = r_{e} \). This means that the Planck horizon is coincident with the stretched horizon and exists exactly at the Schwarzschild radius. Thus, we can interpret open strings in the Callan-Maldacena model as open strings whose ends attach to the Planck horizon. Since the strings can not enter the Planck solid region, strings only exist outside of the Planck horizon in the near extremal black hole case. This fact suggests that a D-brane as the condensed state of strings exists at the Planck horizon, and there is no Hawking radiation since this D-brane is in the ground state. It is expected that we can reproduce the entropy of this black hole by counting the number of the ground states which have the same charges. Therefore, the Planck solid model is not in conflict with the Callan-Maldacena model.

\section{Summary and discussion}
\label{sec:Summary}

In this paper, we have presented an intuitive scenario for the time evolution of the black hole as a possible solution for the information loss problem and the singularity problem, based on some assumptions. In this scenario, a Planck solid is created by a first order phase transition inside a star, and its information and energy are sent outside of its surface. Then, the star is supported by degeneracy pressure of the Planck solid, and the black hole radiation including the information is emitted from its surface. From the difference between this radiation and the thermal Hawking radiation, we can pick up information, and the information loss problem is resolved. Moreover, since the Planck solid has no energy, there is no singularity in the black hole. As time passes, the Planck solid region at the center of the black hole disappears, and the surrounding strings collapse to the thermal radiation. As long as we assume that causality is satisfied, no singularity appears at any time, including a naked singularity, and the singularity problem does not exist. In addition to solving these two important problems, our model gives a physical explanation of the cutoff which is introduced in the brick wall or membrane model.

In this scenario, we made four major assumptions. First, we introduced the Planck solid. We must study string theory at high density and the strong coupling region to investigate the validity of this assumption. However, it seems that the existence of the Planck solid is suggested by the fact that the important problems regarding black holes are resolved in our model. Moreover, as we can see from the above discussion about the cutoff, our Planck solid model proposes the existence of a phase transition from the states of strings to the Planck solid state which has no degrees of freedom, while other people introduce an artificial cutoff. It should be emphasized that in our model a physical entity, the Planck solid has been introduced instead of the artificial cutoff. As we saw in \S \ref{sec:Stringbit}, the properties of the Planck solid are satisfied by topological field theories. For the proof of the existence the Planck solid, it is sufficient to establish these theories as an unbroken phase in string theory.

Next, we assumed that strings in the collapsing body are thermalized in response to the factor \( d \tau / dt \). Because we still do not know precisely the behavior of quantum strings in curved space-time, we cannot tell for certain what happens exactly in this situation. But, the thermalization effect suggests the possibility that if we study the behavior of strings in curved space-time, we can formulate the Hawking-Unruh effect using local gravitational interactions of strings, instead of the computation using the time development of the background metric from the beginning. In any case, we must begin with the construction of a consistent theory for the thermalization of strings with this Hawking-Unruh effect.

Then, we supposed that the black hole is in a quasi-equilibrium state at the Hawking temperature in the last stage of gravitational collapse. Although we do not know the precise relation between our model and Hawking's computation of black hole radiation using quantum field theory in a curved background, this assumption is natural since we can reproduce the black hole entropy and its mass in this framework. In addition to the fact that string theory contains quantum gravity, Susskind pointed out that those strings experiencing thermalization effects interact with each other through gravitational scattering \cite{sus4}. Therefore, we may think that our model shares the viewpoint of high energy gravitational scattering of many strings, rather than that of quantum theory in curved space-time. From our viewpoint, if interactions of strings in a curved background satisfy unitarity, the black hole must also satisfy unitarity. Then, we will be able to construct an {\it S}-matrix of black hole. Hence, the information loss problem would be resolved by determining the time development of a black hole by piling up the local interactions of strings including their back-reactions to the metric, instead of giving the global space-time structure from the beginning.

Finally, we have treated the string system as classical matter and used general relativity. Probably, there are enough strings in this system to treat it as classical matter. In the low energy effective theory of the background field in string theory, massless modes are dilaton and antisymmetric tensor in addition to graviton, and the equations of motion become more complex than Einstein equation. But, in the Planck solid region, these modes are not excited. In any case, the background field in this region would be decided only by boundary conditions, so that we can treat it easily.

After all, we must construct the string theory in the high density, strong coupling region, and also construct quantum string theory in curved space-time in order to confirm whether or not the information loss problem and the singularity problem are resolved in our model. On the other hand, the black hole problems represent a viable context for studying these subjects.

\section*{Acknowledgements}

I am grateful to A. Tokura, S. Hirano and Y. Mino for useful discussions, and especially to K. Kikkawa for reading the manuscript and making a number of helpful suggestions,

\end{document}